\documentclass{PoS}

\usepackage{epsfig}
\usepackage{graphicx}
\usepackage{longtable}

\newcommand{\solm}{M$_{\odot}$\ }

\title{Light and shadow in the Galactic Center
\\
{\normalsize On the detection of the relativistic periastron shift of 
star S2 in the Galactic Center} 
}
\ShortTitle{Light and Shadow in the GC}

\author{\speaker{Andreas Eckart$^{1,2}$},  
M. Parsa$^{1,2}$, 
E. Mossoux$^{3}$
B. Shahzamanian$^{1,4}$, 
M. Zajacek$^{1,2}$, 
E. Hosseini$^{1,2}$, 
M. Subroweit$^1$, 
F. Peissker$^1$, 
N. Sabha$^1$, 
M. Valencia-S.$^1$,
C. Straubmeier$^1$, 
V. Karas$^5$, 
S. Britzen$^2$, 
A. Zensus$^2$\\
1) I. Physikalisches Institut der Universit\"at zu K\"oln, Z\"ulpicher Str. 77, D-50937 K\"oln, Germany;
\\
2) Max-Planck-Institut f\"ur Radioastronomie, Auf dem H\"ugel 69, D-53121 Bonn, Germany;
\\
3)Space sciences, Technologies and Astrophysics Research (STAR) Institute, 
Universit\'e de Li\`ege, All\'ee du 6 Ao\^ut, 19c, B\^at B5c, 
4000 Li\`ege, Belgium;
\\
4) Instituto de Astrofisica de Andalucia (CSIC), Glorieta de la  Astronomia s/n, 18008 Granada, Spain
\\
5) Astronomical Institute of the Academy of Sciences Prague, Bocni II 1401/1a, CZ-141 31 Praha 4, Czech Republic
\\
E-mail: \email{eckart@ph1.uni-koeln.de}
}

\abstract{
We report on the nature of  prominent sources of light and shadow in the Galactic Center.
With respect to the Bremsstrahlung X-ray emission of the hot plasma in that region
the Galactic Center casts a 'shadow'.
The 'shadow' is caused by the Circum Nuclear Disk that surrounds SgrA* at a distance of about 1 to 2 parsec.
This detection allows us to do a detailed investigation of the physical properties of  the surroundings of 
the super massive black hole.
Further in, the cluster of high velocity stars orbiting the central super massive black hole SgrA* 
represents an ideal probe for the gravitational potential and the degree of relativity
that one can attribute to this area. Recently, three of the closest stars (S2, S38, and S55/S0-102)
have been used to conduct these investigations. In addition to the black hole mass and distance
a relativistic parameter defined as $Υ=r_s/r_p$ could be derived for star S2. 
The quantity $r_s$ is the Schwarzschild
radius and $r_p$ is the pericenter distance of the orbiting star. 
Here, in this publication, we highlight the robustness and significance of this result.
If one aims at investigating stronger relativistic effects one needs to get closer to SgrA*.
Here, one can use the emission of plasma blobs that orbit SgrA*.
This information can be obtained by modeling lightcurves of bright X-ray flares.
Finally, we comment on the shadow of the SgrA* black hole expected due to 
light bending and boosting in its vicinity.
}

\FullConference{Frontier Research in Astrophysics - III\\
                28 May - 2 June, 2018\\
                Palermo, Italy}

\begin{document}

%\noindent \textbf{Keywords}: 
%Infrared - Spectroscopy - Photometry - X-rays - individual: Sagittarius~A; Galactic center - black hole physics
%\end{abstract}

\section{Introduction}
\label{section:Introduction}

The Galactic Center is a very active region. It is filled with stars, dust, gas at different
temperatures and different ionization states, and an accreting super massive black hole (SMBH)
at the very center of the region, i.e. Sagittarius A* (SgrA*, for an overview see, e.g., Eckart et al. 2017).
Due to the large mass of SgrA*, it amounts to about 4 million solar masses,
the motion and radiation is also subjected to relativistic effects, which can be used to map
out space time in the vicinity of SgrA*.
This leads to a situation in which many competitive sources of radiation and extinction or
dilution act at the same time. Hence, the region can be thought of as an interplay between
light and shadow.
In this brief overview, we begin the discussion at a scale of several parsecs and investigate the X-ray Bremsstrahlung 
radiation of the plasma, in which SgrA* and the central 
stellar cluster is embedded.
Further in we detect the relativistic motion of the luminous star S2. 
Martins et al. (2008)
confirmed that the star S2 is a main-sequence star of spectral type B0-2.5 V 
with a zero-age main-sequence (ZAMS) mass of 19.5\solm. 
Getting close to the 
last stable orbit around the SMBH we have indications for plasma moving at relativistic speed
and finally there may be the possibility to see the shadow of the SMBH as it bends the light
generated in its accretion zone.

\begin{figure}
	\centering
	\includegraphics[width=\columnwidth]{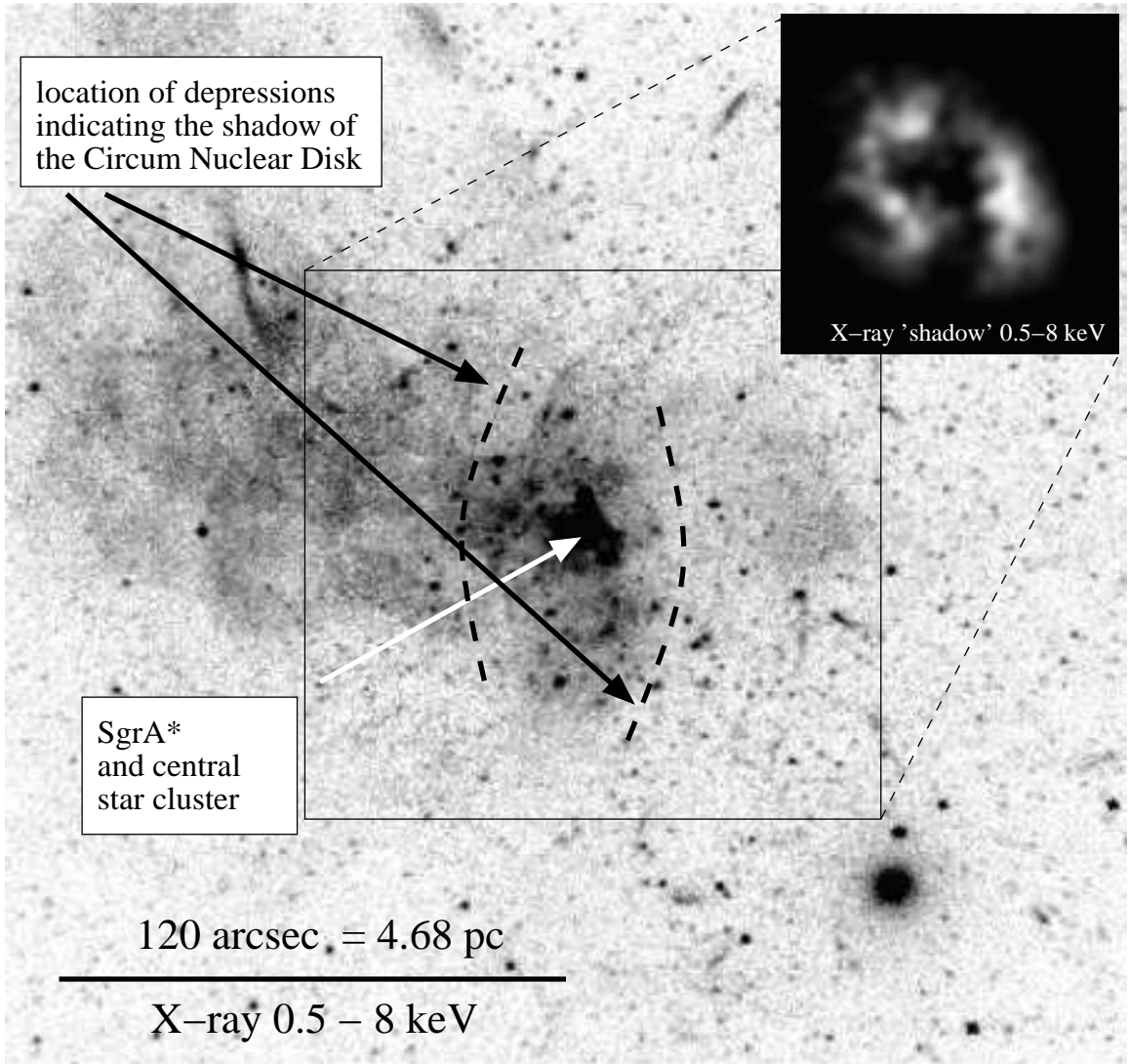}
	\caption{The Chandra image of the X-ray diffuse emission 
at the Galactic Center as shown by Mossoux \& Eckart (2018) and as 
obtained for the energy interval between 0.5 and 8~keV.
The observations have been carried out between 1999 and 2012.
Corresponding to the Chandra angular resolution the pixel size is 
half an arcsecond.
The image intensity represents the count rate. It is shown in 
an inverted logarithmic scale implying that the brightest areas are represented
as the darkest regions.
The location of SgrA* and the central stellar cluster is shown.
The dashed line shows depressions due to the shadowing effect by the CND.
In Mossoux \& Eckart (2018) we explain the algorithm that allowed us to
transform these indications of the shadow into an image of it.
The inset shows the image of the shadow which looks very similar to the 
ALMA image of the CND. Its inner edge is identical to the circumference of
the bright part of the central stellar cluster (see figures in Mossoux \& Eckart 2018).
}
	\label{fig10}
\end{figure}

\subsection{Shadow of the Circum Nuclear Disk}
\label{section:shadow2}

The SgrA complex at the center of the Milky Way consists of several components like
the SMBH SgrA*, the mini-spiral, and the Circum Nuclear Disk (CND).
The source complex can be observed in the radio/sub-millimeter, infrared and X-ray
wavelength domain.
However, until now and as a part of SgrA East, the CND had only been detected in the radio and far 
infrared.
Thanks to the 4.6 Ms of Chandra observations of the Galactic Center region, as described by 
Mossoux \& Eckart (2018), we were now able to detect an X-ray  ''shadow'' of the CND against the
diffuse  X-ray emission of the entire region.
Mossoux \& Eckart (2018) aimed at 
finding out if the CND acts as an absorber for the diffuse X-ray emission
or as a barrier for the plasma close to Sgr A* and the central stellar cluster.
The basis for this investigation are the  ACIS-I and ACIS-S/HETG Chandra data 
covering the time interval between 1999 and 2012.
Subtracting a smooth model of the diffuse X-ray emission from the image, inverting it,
clipping it at zero intensity, and smoothing the result gives an image of the CND with identifiable
subcomponents. The result is rather insensitive to the diffuse X-ray emission model.

The shadow inspired us to define several regions (inside, on, and outside the CND in addition to regions
off the entire structure) for which we extracted spectra. From the spectra we derived temperatures and column
densities. 
The general picture Mossoux \& Eckart (2018) get is that best fits to the spectra are obtained assuming
a two temperature plasma. 
For the CND a single temperature model with 1.8~keV and total local 
column density of 2.3$\times$10$^{22}$cm$^{-2}$ of the hot gas is obtained. 
Inside the CND temperatures of 1~keV and 5~keV are needed. 
Similarly for the outside temperatures of 0.35~keV and 1.3~keV are indicated.
Details of the MCMC fitting results are given in Table 1 by Mossoux \& Eckart (2018).
We find that the plasma at the location of the CND is cooler in general
and may in fact act as a barrier between the hot plasma insight and outside of the CND.

\begin{figure}
	\centering
	\includegraphics[width=\columnwidth]{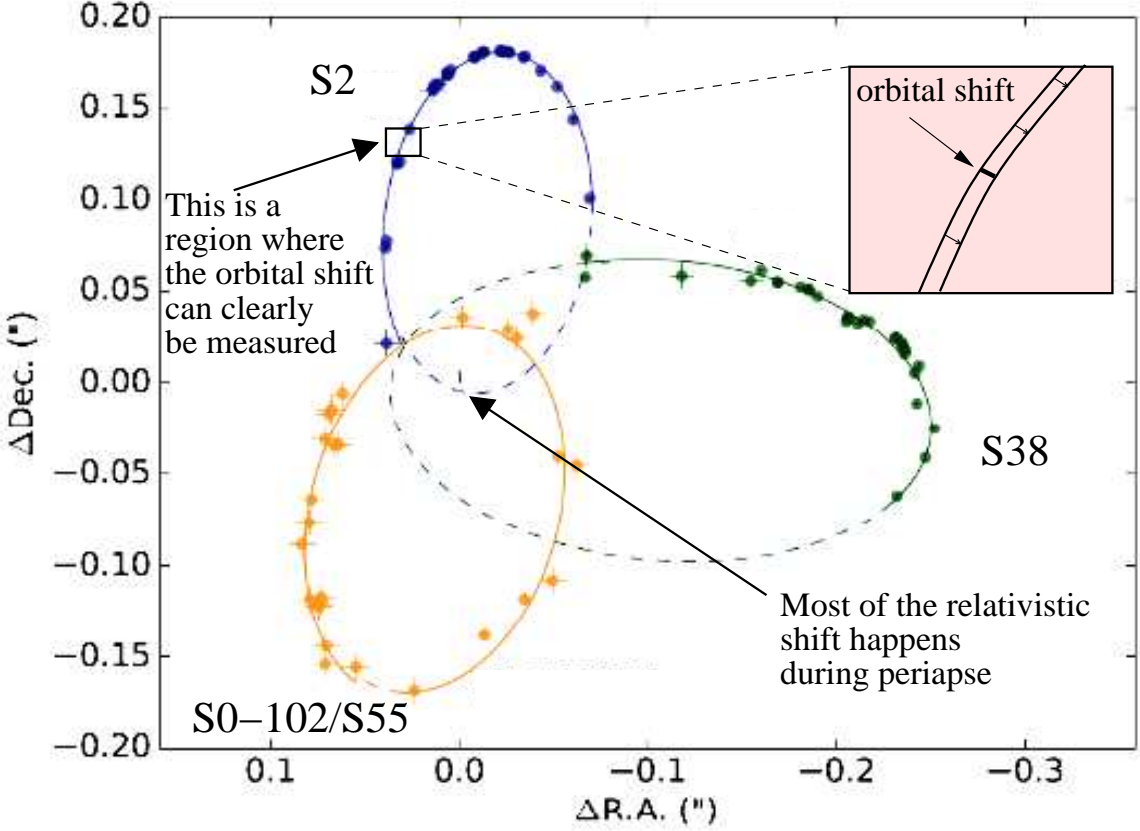}
	\caption{
Orbital fits for the high velocity stars S2, S38, and S0-103/S55 orbiting the Galactic Center
SMBH SgrA* as taken from Parsa et al. (2017). We indicated regions in which the relativistic
periastron shift is imprinted onto the orbit and where the effects can be seen easily 
in comparison to the larger scale orbit. The zoom depicts the (projected) prograde advance 
of the orbit by the amount $\Delta \omega$.
See also text and animation in the ESO press Announcemant ann17051 (http://www.eso.org/public/announcements/ann17051/).
}
	\label{fig10}
\end{figure}

\section{Relativistic motion of star S2}
\label{section:S2}

The S-star cluster that surrounds the supermassive black hole SgrA* in the Galactic center 
is ideally suited to investigate the physics close to this peculiar object.
In particular such a study allows us to perform  dynamical tests of general relativity.
Using positions and radial velocities of three high velocity stars with the 
shortest periods (S2, S38, and S55/S0-102; see Fig.\ref{fig10})
 we derived in Parsa et al. (2017) a black hole mass of 
$M_{BH} = (4.15 \pm 0.13 \pm 0.57) \times 10^6$\solm and a distance to it of
$R_0 = 8.19 \pm 0.11 \pm 0.34~~kpc$. 
The possibility of the detection of faint stars inside the S2 orbit is
discussed in Zajacek \& Tursunov (2018).
The relativistic character of simulated orbits was investigated via  
a first-order post-Newtonian approximation to calculate 
stellar orbits that cover a large range of periapse distances. 
These calculations were used to derive changes in orbital elements 
that were obtained from fits to different sections of the orbits.
Here we used changes in the eccentricity and the semi-major axis obtained 
from fits to the lower and upper part of the
orbits, these are $\Delta e_l/\Delta e_u$ and 
$\Delta a_l/\Delta a_u$. We also calculated the periastron shift $\Delta \omega$ between 
the pre- and post-periapse part of the orbit.
Parsa et al. (2017) could show that these quantities are correlated with 
the relativistic parameter we defined as $Υ=r_s/r_p$.
Here, $r_s$ is the Schwarzschild radius and $r_p$ is the pericenter distance of the orbiting star. 
After establishing these correlations we could use the corresponding data from star S2 
to derive its relativistic parameter and, hence, determine the degree of relativity that is 
shown by its orbit.
For S2 Parsa et al. (2017) were able to derive a value of $Υ=0.00088 \pm 0.00080$.
This value is consistent with the expected value of $Υ=0.00065$ derived for the star S2 using the SgrA* 
black hole mass.
Parsa et al. (2017) argue that this derived value is most likely not dominated by possible perturbing 
influences such as noise on the derived stellar positions, rotation of the field that was imaged at 
different epochs, and possible drifts of the black hole in position.

\begin{figure}
	\centering
	\includegraphics[width=\columnwidth]{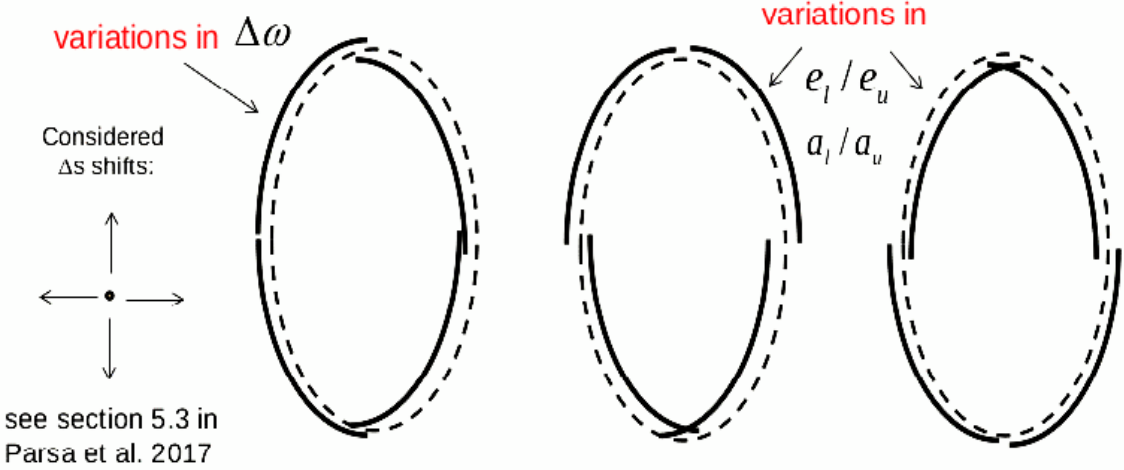}
	\caption{
Sketch depicting the randomization of the shifts for the four
orbital sections by $\Delta s$. As an example of effective variations in 
the periastron shift $\Delta \omega$ and the ratios
$\Delta e_l/\Delta e_u$, $\Delta a_l/\Delta a_u$ between lower and upper 
orbital fits are shown.
Orbital fits to these configurations were used to obtain the 
distributions for the assumption of a fully noise dominated case.
}
	\label{fig20}
\end{figure}

\begin{figure}
	\centering
	\includegraphics[width=\columnwidth]{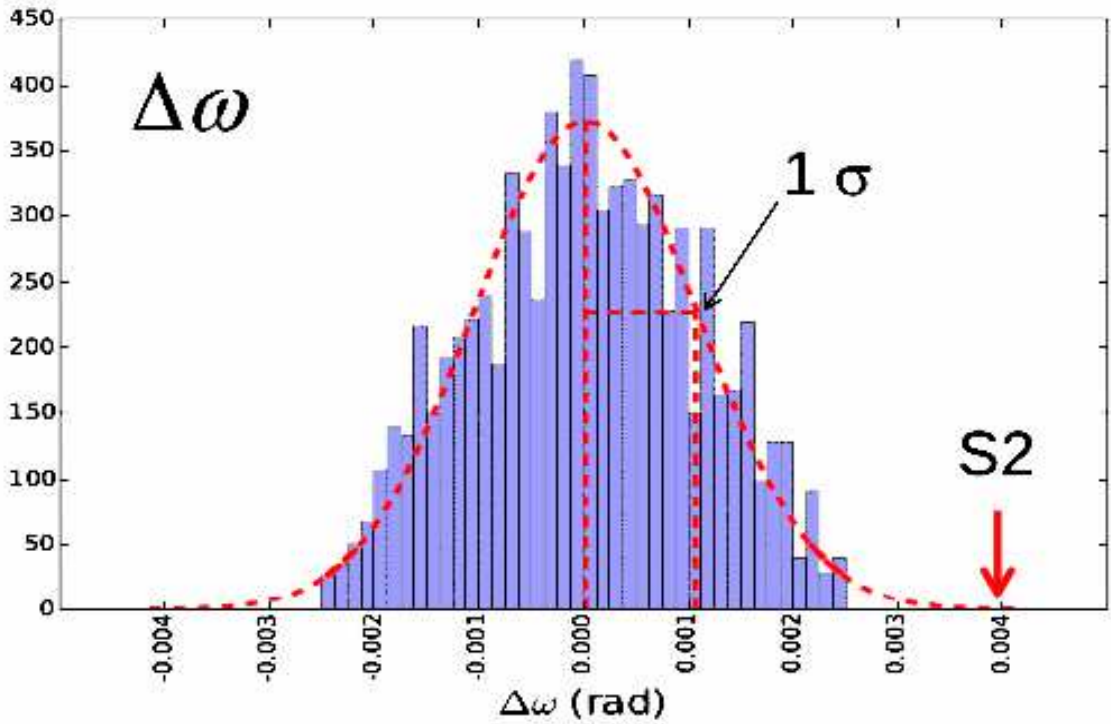}
	\caption{Distribution of the periastron shift $\Delta \omega$ 
obtained from the orbits with randomization of the shifts for the four
orbital sections by $\Delta s$ under the assumption of a 
fully noise dominated Newtonian case.}
	\label{fig30}
\end{figure}

\begin{figure}
	\centering
	\includegraphics[width=\columnwidth]{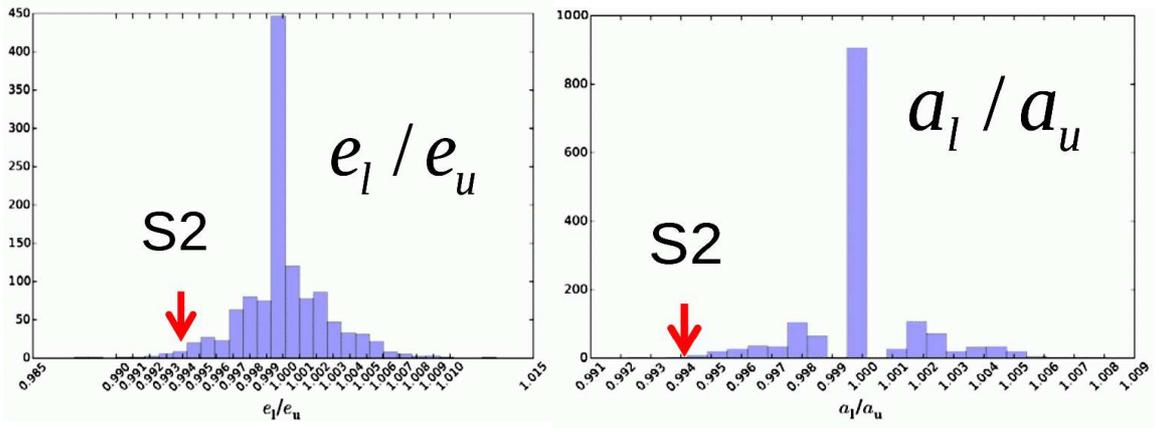}
	\caption{Distribution of 
the ratios $\Delta e_l/\Delta e_u$, $\Delta a_l/\Delta a_u$ 
between lower and upper orbital fits 
obtained from the orbits with randomization of the shifts for the four
orbital sections by $\Delta s$ under the assumption of a 
fully noise dominated Newtonian case.}
	\label{fig40}
\end{figure}

In Parsa et al. (2017) we present a qualitative analysis of the
robustness and significance of the results in chapter~5.3.
Here, we present the results of a numerical calculation of the
significance in comparison to a fully noise dominated scenario.
The uncertainties for the $e$-, and $a$-ratios as well as the $\Delta \omega$
value were obtained by transporting the uncertainties from the 
measurements, via the reference frames to the final statement.
As we used only images in which SgrA* could be detected as well,
for the stars in the central arcsecond
the positional uncertainties are the most important quantities in order
to measure the deviation from a Newtonian orbit.
We assume a noise dominated case and estimate the uncertainties relative to it.
We use the combination of our uncertainty in R.A. direction
(essential in the measurement of the $\Delta \omega$ of the S2 orbit) 
and the data from the literature.
We then find a mean uncertainty of 1.4 mas for an individual position.
Knowing that we (Parsa et al. 2017) have about 5 data points per quarter of the three dimensional 
orbit we derive a positioning uncertainty for each quarter of 
about $\Delta s$ = 0.5~mas (in the projected view of the orbit).
Hence, we randomize the position of each orbital segment 
by placing it at $\Delta s$=0,+0.5,-0.5~mas (see Fig.\ref{fig20})
with respect to the nominal position in the deprojected Newtonian orbit, i.e., less than that 
in the projected view of the orbit.
For each orbit that we generate with this procedure we calculate the 
quantities $\Delta e_u/\Delta e_l$, $\Delta a_u/\Delta a_l$, and 
the periastron shift $\Delta \omega$. The corresponding histograms are 
shown in Figs.\ref{fig30} and \ref{fig40}. 
These diagrams can now be used to determine an
uncertainty $\sigma$ of the distributions and compare those to the values measured off for star S2. 
The $e$- and $a$-ratios and $\Delta \omega$ obtained for star S2 represent at least 3-4$\sigma$ excursions
including the 3D application of $\Delta s$ (see above).
Taking the inclination of S2 of about 137$^o$ into account the 3-4$\sigma$ excursions turn into a 4-6$\sigma$ result.
Parsa et al. (2017) shows that the inclusion of a possible small proper motion of SgrA*
with respect to the stellar cluster does not change the result on the relativistic periastron shift
(see in particular Tab.8 in Parsa et al. 2017). Hence, it also does not effect our evaluation of the result
with respect to the noise dominated Newtonian case.

\subsection{Relativistic motion of plasma blobs}
\label{section:blobs}

Karssen et al. (2017) could show that the lightcurves of bright X-ray SgrA* flares
have the exact shape one expects from plasma blobs orbiting the SMBH close to the
last stable orbit. This is true for the four bright flares discussed by Karssen et al. (2017)
and the recent bright flare reported by Ponti et al. (2017) (see also Eckart et al. 2018).
Since the bright X-ray flares are correlated with synchronous NIR flares 
(see e.g. Eckart et al. 2012)
the finding by
Karssen et al. (2017) probably also holds for the NIR domain. 
However, in the X-ray domain, flares occur at a low rate than in the NIR. 
Therefore, it is less likely that faint  X-ray
flares overlap the bright flare events resulting in a clearer agreement with 
lightcurves expected from orbiting blobs. In the NIR flares are much more frequent in it is less
likely to obtain a clean single blob light curve undisturbed by secondary flare events.
The assumption that goes in is that the blobs stay stable for a substantial fraction of 
a single orbit.
The characteristics of these theoretical light curves is that they all 
show a shoulder due to 
lensing amplification on the ascending part of the boosting dominated light curve
(The boosting sets in before the lensing occurs). This can be found in the observed lighcurves of all bright
X-ray flares as well. Reassuring is also that by fitting the flare curves scale free (in gravitational time scales $GM/c^3$) with the theoretical light curves and then scaling them by introducing the observed 
flare time in seconds  one obtains the mass of SgrA*. This can be understood by the fact that the 
black hole mass scales linearly with the black hole 
(or rather the event horizon) size,i.e., with the orbital time scale 
and, therefore, the flare time scale in our model.
Hence, being able to explain the X-ray flare shapes via the assumption of orbiting matter 
is another indication for the high degree of relativity in the vicinity of SgrA*.

\subsection{Shadow of the supermassive black hole}
\label{section:shadow1}

One of the goals of the Event Horizon Telescope project (EHT) is to get a high
angular resolution image at mm-wavelengths of SgrA* in order to measure the
so called shadow of the SMBH. 
This shadow results from the following effect:
Luminous plasma is orbiting the SMBH.
Due to the bending of light by the SMBH there is the possibility
that in addition to a bright Doppler boosted part of the source 
there may appear a dark region in the overall structure (see Falcke et al. 2000).
The expected size of that region  is of the order of 30-50$\mu$as.
However, if one is able to see it or not will critically depend on the actual observational 
and geometrical situation (The shadow is also not a feature uniquely associated to Black holes only - see detailed discussion in Eckart et al. 2017). 
In order to see this shadow one needs to have a special orientation of the system, 
the accretion must be well ordered and not chaotic, and it would be best if there 
is no jet that may disturb the impression one gets from the orbiting material.
Another problem may also result from the fact that the scattering screen
along the line of sight just becomes transparent around a wavelength of 1~mm.
Hence, it may be that multiple simultaneous images (like the speckle effect known
from observations at optical and infrared wavelengths) may occur. 
All of these effects and influences will make it difficult to get a clear view 
of the immediate environment of SgrA*.

The first EHT results were recently published by Ru-Sen Lu et al. (2018).
The authors report on a detection of intrinsic source structure at about 
3 Schwarzschild radii, i.e., about 30$\mu$as and therefore on the expected 
size scale for the shadow of the black hole.
Their Fig.5 shows possible image structures that one would expect from 
based on the current amplitude and closure phase measurements.
The variations in the data and the appearance of the models are also 
consistent with the expectation of a shadow, however, detailed images that 
would show that disk structure due to orbiting matter with indications of a 
shadow have not yet been published.

\begin{figure}
	\centering
	\includegraphics[width=\columnwidth]{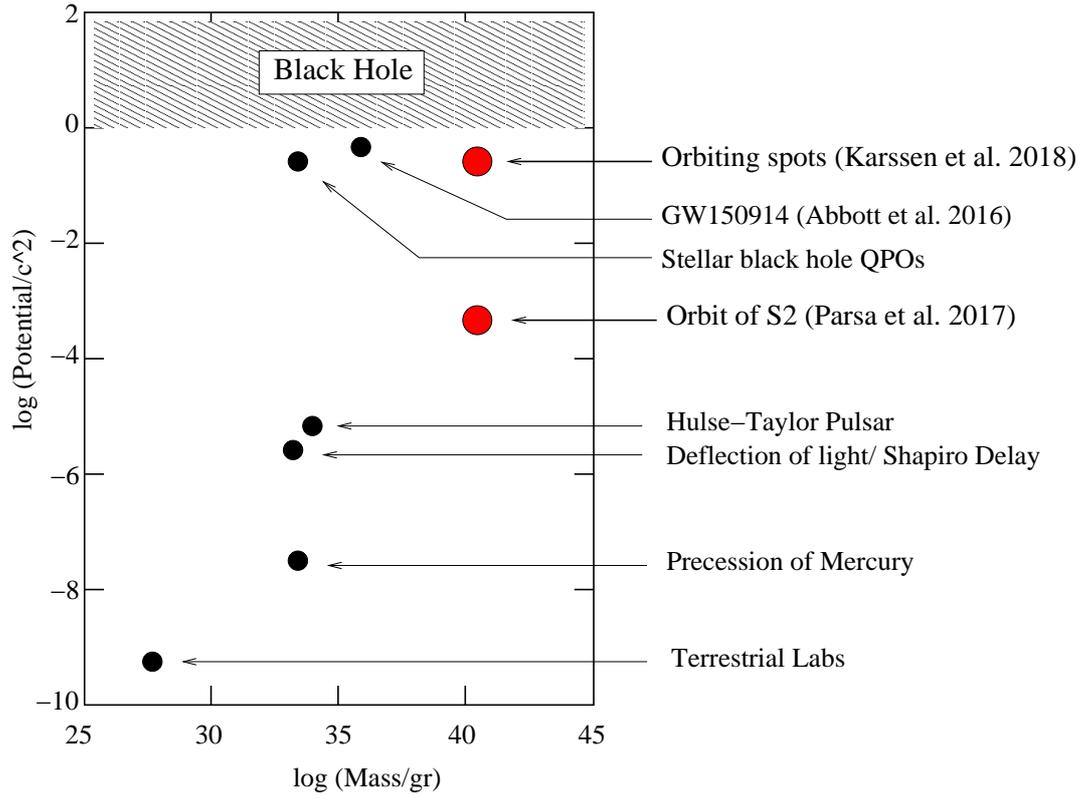}
	\caption{
The gravitational potential that is measured by a number of tests of gravity 
presented in comparison to the mass responsible for the potential (based on Psaltis 2004).
We have selected terrestrial labs, the precession of Mercury, light deflection and the 
Shapiro	delay in the solar system, the Hulse-Taylor pulsar, QSO QPOs, and the gravitational
wave detection GW150914 (Abbott et al. 2016). In addition we show the 
results for the Galactic Center S2 orbit presented by Parsa et al. (2017) and the
X-ray lightcurve fitting presented by Karssen et al. (2017).
}
\label{fig50}
\end{figure}

\section{Summary and Conclusion}
\label{section:summary}

Luminous stars and plasma blobs can be used to map the space time close to SgrA*.
While the presence of the plasma blobs close to the last stable orbit needs to be
inferred from the shape of the light curves the stars can be observed directly.
The plasma blobs would be part of the structure surrounding the SMBH.
Due to light bending effects that structure is potentially showing a shadow of the black hole.

With Parsa et al. (2017) we used three stars to derive the mass and distance
of SgrA* in a Newtonian and post-Newtonian solution.
A new and simple method that compares properties of Newtonian fits
to the lower, upper, pre-, and post-periapse parts of the orbits
allows us to determine the degree of relativity.
For the high velocity star S2 the values for the 
$e-$ and $a-$ratios as well as $\Delta \omega$ value
lie close to the values expected for S2 and the SgrA* mass (see Iorio 2017).
Here we give an analysis of the uncertainties for these quantities 
with respect to a noise dominated Newtonian situation. In this case the deviations
are significant and the S2 values for
the e- and a-ratios and $\Delta \omega$ represent a $\ge$4~$\sigma$ result,
and it appears to be highly unlikely that the common effect seen 
in all three quantities originates from a purely noise dominated scenario.
Accepting this result, S2 is the first star with a resolvable orbit
around a SMBH for which a test for relativity can be performed.
In the near future these results will become more significant using
interferometric data as obtained by GRAVITY at the VLTI
(e.g., Gravity Collaboration 2017, M\'erand et al. 2017).

In addition we report on the 'shadow' the Galactic Center casts with respect to the 
X-ray Bremsstrahlung emission of the plasma in the local for- and background.
The 'shadow' is caused by 
The Circum Nuclear Disk that surrounds SgrA* casts a 
'shadow' that allows us to effectively subdivide the hot nuclear plasma into
areas inside, at, and outside the CND. Two temperature models and estimates of the
column density give us a clear picture of the physical conditions of the plasma.
The CND appears to act as a blocking agent that keeps the hot plasma inside the central stellar cluster
and close to the super massive black hole SgrA*.
 
The probes of the gravitational potential as they have been presented by
Parsa et al. (2017) and Karssen et al. (2017) and summarized in the present publication
can be compared to other tests in Fig.~\ref{fig50}. This comparison is based on
Fig.6 in Psaltis et al. (2004; see also Fig.1 in Hees et al. 2017, and
Figs.4 and 5 in Eckart et al. 2010).
This figure shows that the gravitational potential can now be probed over 9 orders of
magnitude for masses covering more than 10 orders of magnitude.
These results can still be improved with currently running interferometry experiments 
in the radio mm-wavelength and infrared regime.
\\
\\
{\bf Acknowledgements:}
We received funding from the European Union Seventh Framework Program (FP7/2007-2013)
under grant agreement No. 312789 - Strong gravity: Probing Strong Gravity by Black
Holes Across the Range of Masses. This work was supported in part by the Deutsche
Forschungsgemeinschaft (DFG) via the Cologne Bonn Graduate School (BCGS), the Max
Planck Society through the International Max Planck Research School (IMPRS) for
Astronomy and Astrophysics, as well as special funds through the University of
Cologne and SFB 956 - Conditions  and Impact of Star Formation. 
E. Hosseini is members of the IMPRS. 
We thank the Czech Science Foundation - DFG collaboration (No. 13-00070J) - and the
German Academic Exchange Service (DAAD) for support under COPRAG2015 (No.57147386).
The result on S2 has also been presented at the
'Stellar Dynamics in Galactic Nuclei' workshop,
held at the Institute for Advanced Study, in Princeton, NJ,
between November 29 and December 1, 2017,
and at the 
'Second Annual Black Hole Initiative Conference on Black Holes'
(Harvard University),
held at the Sheraton Commander Hotel, 16 Garden Street, Cambridge, MA,
Wednesday, May 9 through Friday, May 11, 2018.

\end{document}